\documentclass[reprint,aps,pra,
amsmath,amssymb,
twocolumn,superscriptaddress
]{revtex4-2}
\usepackage{revquantum}
\usepackage[super]{nth}
\usepackage{graphicx}
\usepackage{bm}
\usepackage{dsfont}

\usepackage{hyperref}

\newcounter{temprom}
\begin{document}
\preprint{APS}
\title{\textbf{Physically Motivated Ansatz for Open Fermionic Systems on Quantum Computer
}}
\author{Yi Liu}
\affiliation{State Key Laboratory of Precision and Intelligent
Chemistry, University of Science and Technology of China,
Hefei, Anhui 230026, China}
\author{Xiaopeng Li}
\affiliation{State Key Laboratory of Precision and Intelligent
Chemistry, University of Science and Technology of China,
Hefei, Anhui 230026, China}
\author{Zhen Liu}
\affiliation{State Key Laboratory of Precision and Intelligent
Chemistry, University of Science and Technology of China,
Hefei, Anhui 230026, China}
\author{Zhenyu Li}%
\email{zyli@ustc.edu.cn}
\affiliation{State Key Laboratory of Precision and Intelligent
Chemistry, University of Science and Technology of China,
Hefei, Anhui 230026, China}
\affiliation{Hefei National Laboratory, University of Science and Technology of China, Hefei,
Anhui 230088, China}

\begin{abstract}
Variational quantum algorithms offer a promising route to solving the Lindblad equation for open quantum systems, analogous to their application to the Schrödinger equation in closed systems. However, ansatz design for open systems is more challenging and it still relies on heuristic hardware-efficient ansatzes (HEA) susceptible to barren plateaus. In this study, we introduce a physically motivated ansatz, NE-UCC. Numerical simulations demonstrate that it can be reliably converged to steady states, reducing the infidelity by up to ten orders of magnitude compared to HEA. Furthermore, NE-UCC enables convenient search for excited modes with specific symmetries. 
\end{abstract}

\maketitle


\section{Introduction}
Open quantum systems are central to diverse fields, including quantum hardware~\cite{quanth}, quantum transport~\cite{revlind1,revlind2}, and biochemical processes~\cite{bio1}. Unlike closed systems governed by Hamiltonian dynamics, open systems relax under a non-Hermitian Liouville superoperator $\mathcal{L}$ (Liouvillian)~\cite{open}, as exemplified by the Lindblad master equation~\cite{lindblad0,lindblad1,lindblad2}. The Liouvillian simultaneously encodes system–reservoir coupling, interparticle interactions, and finite thermodynamic biases. Combined with the intrinsic nonequilibrium nature and the exponential growth of the Hilbert space, these features make the determination of non-equilibrium steady states (NESS) a formidable theoretical challenge. Analytically solvable many-body NESS remain rare~\cite{thirdq,sf,xxz1,xxz2}. While they offer powerful strategies to circumvent the exponential barrier, tensor network and quantum Monte Carlo methods~\cite{genal1,genal2,genal3,genal4,genal5,genal6,genal7,genal8} are respectively limited by the area law and the sign problem.

Quantum computation, with its intrinsic exponential expressive power and the ability to generate global entanglement, provides a promising way to address this challenge. Various quantum algorithms have been proposed for open quantum systems~\cite{zhengxiao,hu1,hu2,choi1,choi2,pure1,yuan1,yuan2,yuan3,dvqe1,dvqe2,q2lind}. One can either implement nonunitary evolution via unitary dilation of completely positive maps~\cite{hu1,hu2,choi1,choi2} or employ parameterized quantum circuits to represent a pure-state trajectory~\cite{pure1,yuan1,yuan2,yuan3}. To directly target NESS without accumulating time-evolution errors, it is more convenient to variationally minimize the corresponding density matrix~\cite{dvqe1,dvqe2,q2lind}, similar to wavefunction optimization in the variational quantum eigensolver (VQE) algorithm for closed systems. 

The performance of a variational quantum algorithm is mainly determined by the ansatz design. In VQE simulations, ansätze with solid physical foundation, such as unitary coupled cluster (UCC)~\cite{vqe1} and its variant ADAPT~\cite{vqe2} have been widely used, showing the ability to attain a high accuracy~\cite{ucc1,ucc2,ucc3,ucc4}. Unfortunately, owing to limited insight into the physical structure of the Liouvillian, quantum algorithms for open systems continue to rely on heuristic hardware-efficient ansätze (HEA)~\cite{hea} — an approach plagued by the barren plateau problem~\cite{barren}, where gradients vanish exponentially with system size, making optimization extremely challenging. At the same time, without physical insight, the vast space of possible HEA configurations makes identifying a relatively good architecture highly nontrivial. 


In this study, by introducing a weak Liouvillian symmetry~\cite{symmetry1,sfci,sfcc,genal1} into the ansatz design, we successfully extend the UCC-based VQE framework to open fermionic systems, which we name NE-UCC. It is fully compatible with the ADAPT-VQE framework~\cite{vqe2}, enabling a flexible trade-off between accuracy and computational cost. We use even-parity Lindblad master equations with a nonrelativistic electronic Hamiltonian and fermionic jump operators, which are widely applied to nonequilibrium transport and impurity problems~\cite{sf,sf1,sfci,sfcc,genal0,genal1,genal2,genal3,genal4},  as examples to test the performance of NE-UCC. Since the variational search space is substantially reduced from the full super-Fock space to the physically relevant symmetry-preserving sectors, high accuracy is reached for these systems where previous HEA ansätze fail to even give a qualitative description. Moreover, with symmetry introduced, it is very convenient to study excited modes with specific symmetries. NE-UCC thus offers a robust and scalable route for simulating open fermionic quantum systems on quantum computers.

\section{theory}
We first briefly review the superfermion representation~\cite{sf,genal1}, which can be employed to describe Liouvillians obeying the statistics of identical particles. The superfermion representation offers an alternative encoding scheme for Liouville operators: it allows us to conveniently vectorize the density matrix and recast the Lindblad master equation into a non-Hermitian Schrödinger equation defined on the doubled Fock space. More importantly, within the superfermion formalism, we can exploit the weak symmetries of the Liouvillian~\cite{symmetry1} to construct unitary coupled-cluster ansätze tailored for open fermionic systems.
\subsection{Superfermion formalism and weak symmetry}
We consider the Lindblad master equation for a fermionic system
\begin{align}
    i\frac{d}{dt}\rho=\mathcal{L}\rho,
    \label{eq:lind0}
\end{align}
where $\rho$ is the density matrix and we set \(\hbar=k_{\text {B}}=-e=1\). The right hand side of the equation is 
\begin{align}
\label{eq:lind1}
    \mathcal{L}\rho=&[ H,\rho]+i\sum_{i}(2 L_i\rho  L_i^\dagger-\{ L_i^\dagger  L_i,\rho\}),
\end{align}
where
\(H(a,a^\dagger)=\sum_{p,q} f_{p,q}a_p^\dagger a_q+\sum_{p,q,r,s} v_{pqrs}\, a_p^\dagger a_q^\dagger a_s a_r\)
is a typical even-order non-relativistic Hamiltonian, whereas the jump operators take the form \(L_i=\sqrt{\Gamma_i}\,a\) or \(L_i=\sqrt{\Gamma_i}\,a^\dagger\).
The Liouvillian admits a matrix representation on a doubled Hilbert space \(\mathcal{H}\otimes \mathcal{H}\), typically achieved via a vectorization mapping \(A\rho B\rightarrow A \otimes B^T|\rho\rangle\)\cite{vec1,vec2,vec3}. 

In this study, however, we adopt a vectorization scheme based on the superfermion formalism~\cite{sf,sf1,sfci,sfcc} to naturally incorporate fermionic statistics. This approach establishes an isomorphism between the operator space and the doubled Fock space by treating the basis operators as state vectors via the mapping \(|m\rangle\langle n |\rightarrow|m\rangle|\tilde{n}\rangle\).
Then, we introduce a maximally entangled state, termed the left vacuum
\begin{align}
    |I\rangle=\sum_{n} P_{n}|n\rangle\otimes{|\tilde{n}\rangle},
\end{align}
with \(|P_{n}|=1\). The density state vector can be defined as
\begin{align}
    |\rho\rangle:=\rho\otimes {\tilde{\mathds{1}}}|I\rangle =\sum_{mn}\rho_{mn}P_{n}|m\rangle \otimes|\tilde{n}\rangle,
\end{align}
with the normalization condition \(\operatorname{Tr}\rho=1\) being translated to 
\begin{equation}
    \begin{aligned}
    \langle I|\rho\rangle&=(\sum_{j}\langle j|\otimes\langle \tilde{j}|P_j^*)(\sum_{mn}\rho_{mn}P_n|m\rangle \otimes|\tilde{n}\rangle)\\
    &=\sum_{j}\rho_{jj}P_j^*P_j=1\\
\end{aligned}
\end{equation}
and, analogously, the average value of an observable operator \({A}\) is given by
\begin{align}
    \langle {A}\rangle =\operatorname{Tr}({A}\rho ):=\frac{\langle I|A\otimes{\tilde{\mathds{1}}}|\rho\rangle}{\langle I|\rho\rangle}
\end{align}
The physical requirement of trace preservation,
\begin{align}
    0=i\frac{d}{dt}\langle I|\rho(t)\rangle=\langle I|{\mathcal{L}}|\rho(t)\rangle,
\end{align}
implies that the left vacuum is a left zero-eigenstate of the Liouvillian, i.e.,
\begin{align}
    \langle I|\mathcal{L}=0.
\end{align}

A key aspect of the superfermion formalism is the tilde conjugation, which maps physical fermion operators (\(a,a^\dagger\)) to auxiliary fermion operators (\(\tilde{a},\tilde{a}^\dagger\)). By choosing the phase factors as \(P_{\{n\}}=(-i)^{n_1+n_2+\dots +n_N}\) within Fock space, the left vacuum reads
\begin{align}
        |I\rangle=\sum_{\{n\}} P_{\{n\}}|n_1,\dots,n_N\rangle\otimes\widetilde{|n_1,\dots,n_N\rangle},
\end{align}
and the tilde conjugation relations follow:
\begin{equation}
    \begin{aligned}
    a_j|I\rangle&=-i\tilde{a}_j^\dagger|I\rangle \quad \overset{\text{adjoint}}{\longleftrightarrow} \quad \langle I|a_j^\dagger=\langle I|i\tilde{a}_j\\
    a_j^\dagger|I\rangle&=-i\tilde{a}_j|I\rangle \quad \overset{\text{adjoint}}{\longleftrightarrow} \quad \langle I|a_j=\langle I|i\tilde{a}_j^\dagger.
    \label{eq:adj0}
\end{aligned}
\end{equation}
Then we left-multiply both sides of Eq.(\ref{eq:lind0}) by the left vacuum \(|I\rangle\), using the above relations in Eq.(\ref{eq:adj0}) and the fact that the original density matrix \(\rho(a,a^\dagger)\) commutes with all tilde operators, then we can cast the Lindblad equation into the Schrödinger equation form. 
\begin{equation}
i\frac{d}{dt}|\rho\rangle=\hat{\mathcal{L}}|\rho\rangle
\end{equation}
where
    \begin{align}
    \hat{\mathcal{L}}={H}-\tilde{H}-i\sum_k \Pi_k,
        \label{eq:lop}
    \end{align}
and the non-Hermitian dissipators \(\Pi_k\) are given by
\begin{align}
    \Pi_k={L}_k^\dagger {L}_k+\tilde{L}^\dagger_k \tilde{L}_k +2i{L}_k\tilde{L}_k.
\end{align}

Due to the formal symmetry between the original operators and their tilde counterparts, it is straightforward to verify that the particle number difference operator, defined as
\begin{align}
    {N}_a-{N}_{\tilde{a}} =\sum_i(a_i^\dagger a_i -\tilde{a}_i^\dagger \tilde{a}_i),
\end{align}
commutes with the Liouvillian
\begin{align}
    [{N}_a-{N}_{\tilde{a}},\hat{\mathcal{L}}]=0.
\end{align}
Notice that, the physical particle number operator \(N_a\) generally commutes only with the Hamiltonian, but not with the jump operators. Consequently, the particle number difference serves as the generator of unitary weak symmetry of the Liouvillian~\cite{symmetry1}, \(\mathcal{U}(\theta)=e^{i\theta N_a}\otimes e^{-i\theta N_{\tilde{a}}}\).

\subsection{NE-UCC ansatz inspired by weak symmetry}
NESS is a fixed point of the evolution \(\frac{d}{dt}\rho=0\) and it corresponds to the diagonalization problem of finding the zero-eigenvalue eigenstate of the Liouvillian \(\hat{\mathcal{L}}|\rho\rangle=0\). Since \({\hat{\mathcal{L}}}^\dagger{\hat{\mathcal{L}}}\ge 0\), it can be cast as a variational problem of minimizing~\cite{dvqe1,dvqe2}
\begin{align}
    \mathcal{E}(\vec{\theta})=\langle \rho(\vec{\theta})|{\hat{\mathcal{L}}}^\dagger{\hat{\mathcal{L}}}|\rho(\vec{\theta})\rangle
    \label{eq:costfun}
\end{align}
with respect to  parameters $\vec{\theta}$ of the circuit to prepare the density state \(|\rho(\vec{\theta})\rangle\). Now, we design a quantum circuit that restricts \(|\rho(\vec{\theta})\rangle\) to the physical subspace. Leveraging the intrinsic structure of the superfermion formalism, we note that the left vacuum (which is the left zero-eigenstate) and the steady state both reside in the subspace defined by \({N}_a-{N}_{\tilde{a}} =0\).
This subspace is spanned by applying excitation and de-excitation operators that conserve \({N}_a-{N}_{\tilde{a}}\) and the density state can thus be expanded as \(|\rho\rangle=(1+\sum t^{ij}a_i^\dagger \tilde a_j^\dagger +\sum t_{ij}a_i \tilde a_j+\sum t_{jl}^{ik}a_i^\dagger a_j \tilde a_k^\dagger\tilde a_l +\dots)|\rho_0\rangle\), where \(|\rho_0\rangle\) denotes a state vector residing within the subspace satisfying \(N_a - N_{\tilde{a}} = 0\). Such an expansion can be viewed as the full configuration interaction (FCI) ansatz for the NESS~\cite{sfci,sfcc}. 

To make truncation, a UCC style framework is proposed.
Analogous to standard truncated coupled cluster methods, the cluster operator comprising single (\(T_1\)) and double (\(T_2\)) excitations over \(N\) orbitals is defined as:
\begin{align}
    T_1=\{a_p^\dagger \tilde{a}_q^\dagger-a_q \tilde{a}_p \mid p,q\leq N\}
\end{align}
\begin{equation}
    \begin{aligned}
    T_2=\{&a_p^\dagger a_q^\dagger \tilde{a}_r^\dagger  \tilde{a}_s^\dagger- \tilde{a}_s  \tilde{a}_r a_q a_p,\\
    &a_p^\dagger a_q \tilde{a}_r^\dagger \tilde{a}_s - a_s^\dagger a_r \tilde{a}_q^\dagger \tilde{a}_p \mid p,q,r,s\leq N\}.
\end{aligned}
\end{equation}
Notice that, unlike UCC for closed systems, the density matrix, represented as a state vector \(|\rho\rangle\), is generally a complex-valued state.
Consider the general case where the derivative of the cost functional \(\mathcal{E}[|\phi(z)\rangle]=\langle\phi_0|e^{{G}(z)^\dagger} E e^{ {G}(z)}|\phi_0\rangle\) with respect to an anti-Hermitian generator operator \({G}(z)=x({\tau}-{\tau}^\dagger)+y i({\tau}+{\tau}^\dagger)\) with complex parameter \(z=x+iy\):
\begin{align}
    \frac{\partial \mathcal{E}}{\partial x}=\langle\phi(z)|[{E},{\tau}-{\tau}^\dagger]|\phi(z)\rangle,
\end{align}
\begin{align}
    \frac{\partial \mathcal{E}}{\partial y}=\langle\phi(z)|[{E},i({\tau}+{\tau}^\dagger)]|\phi(z)\rangle.
    \label{eq:imacse}
\end{align}
Therefore, in general cases, in addition to  \(({\tau}-{\tau}^\dagger)\), operators of the form \(i({\tau}+{\tau}^\dagger)\) should also be introduced~\cite{compl,eom-ucc-fanyi}.  These complementary operators are defined as
\begin{align}
    T_1^c=\{i(a_p^\dagger \tilde{a}_q^\dagger+a_q \tilde{a}_p) \mid p,q\leq N\}
\end{align}
\begin{equation}
    \begin{aligned}
    T_2^c=\{&i(a_p^\dagger a_q^\dagger \tilde{a}_r^\dagger  \tilde{a}_s^\dagger+ \tilde{a}_s  \tilde{a}_r a_q a_p),\\
    &i(a_p^\dagger a_q \tilde{a}_r^\dagger \tilde{a}_s + a_s^\dagger a_r \tilde{a}_q^\dagger \tilde{a}_p) \mid p,q,r,s\leq N\}.
\end{aligned}
\end{equation}
We adopt a second-order truncation, referred to as NE-UCCSD, for which the complete operator pool \(\mathcal{O}_{\text{SD}}\) is given by:
\begin{align}
    \mathcal{O}_{\text{SD}}=T_1 \cup T_1^c \cup T_2 \cup T_2^c .
    \label{eq:pool}
\end{align}
The corresponding NE-UCCSD wave function under the first-order Trotter–Suzuki approximation follows
\begin{align}
    |\rho_\infty\rangle=\prod_{i=1}^{\dim\mathcal{O}_{\text{SD}}} e^{\theta_i{t}_i}|\rho_0\rangle,
\end{align}
where \({t}_i \in \mathcal{O}_{\text{SD}}\) and \(|\rho_0\rangle\) is a reference state.

In this work, to simplify the encoding of Liouvillian operators and the construction of NE-UCC quantum circuits, we project the super-Fock space onto a 2N-dimensional Fock space, \(|n\rangle \otimes |m\rangle \to |n,m\rangle\), i.e., \(\tilde{a}_i = a_{i+N}\). Correspondingly, the left vacuum is further written as
\begin{align}
    |I\rangle = \sum_{\{n_i\}} (-i)^{(\sum_{i=1}^N n_i) \bmod 2} |\{n_i\},\{\tilde{n}_i\}\rangle
\end{align}
Under this setup, the tilde rule in Eq.~(\ref{eq:adj0}) still holds. Then all tilde operators can be expressed in terms of canonical fermionic operators (\(\bm a,\bm a^\dagger\)). Therefore, standard UCC compilation strategies~\cite{cmpl1} can be employed for our NE-UCC method.

\section{Numerical simulation}
To demonstrate the performance of the algorithm, we apply the NE-UCC ansatz to a boundary-driven XXZ transport model~\cite{xxz1,xxz2} and an interacting three-site quantum heat engine~\cite{genal3}, both of which feature strong system-bath coupling and strong interactions simultaneously. In the NESS search algorithm, the initial state \(|\rho_0\rangle\) can be explicitly constructed using the analytical formula \(|\rho_{0}\rangle = \prod_{i=1}^{n} c_i \tilde{c}_i |0\rangle \otimes |\tilde{0}\rangle\), where \(c\) and \(\tilde{c}\) are non-canonical fermionic operators derived from numerically tractable analytical expressions of quadratic Liouvillian, see Appendix.~\ref{apd:sfscf}. Furthermore, by virtue of the symmetry-protected properties of NE-UCC, we demonstrate its capability to search for excited eigenmodes within a designated subspace. Numerical simulations of quantum circuits are run on the state vector simulator of our homemade quantum computing platform, Q\(^{2}\)Chemistry~\cite{q2chem}. We adopt the BFGS algorithm \cite{bfgs} implemented within SciPy \cite{scipy} as our classical optimizer, and utilize reverse-mode automatic differentiation \cite{revmod} (commonly referred to as backpropagation) to supply first-order gradient information for this optimizer.

\subsection{XXZ boundary driven model}
The \(n\)-site boundary driven XXZ chain follows the Lindblad equation with Hamiltonian
\begin{align}
    H_\text{XXZ}=\sum_{i=1}^{n-1}(J\sigma_i^x\sigma_{i+1}^x+J\sigma_i^y\sigma_{i+1}^y+\Delta\sigma_i^z\sigma_{i+1}^z ).
\end{align}
The system features symmetric Lindblad driving defined by
\begin{align}
    L_{1,2}&=\sqrt{\frac{1}{2}\varepsilon(1\pm \mu)}\sigma_1^{\pm}=\sqrt{\gamma^{\pm}}\sigma_1^{\pm}\\
    L_{3,4}&=\sqrt{\frac{1}{2}\varepsilon(1\mp \mu)}\sigma_n^{\pm}=\sqrt{\gamma^{\mp}}\sigma_n^{\pm}
\end{align}
where \(\sigma_k^{\pm}=\frac{1}{2}(\sigma_k^x \pm i\sigma_k^y)\) and \(\sigma_k^x\), \(\sigma_k^y\), \(\sigma_k^z\) are Pauli operators acting on site \(k\). This model is equivalent to an interacting spinless fermion model via the Jordan–Wigner (JW) transformation. Therefore, the superfermion formalism is equally applicable; detailed derivations are given in the Appendix.~\ref{apd:xxzmd}.

Here, we consider a 12-qubit system. After removing redundant operators, the NE-UCCSD ansatz contains 2322 variational parameters. For comparison, we adopt two HEA ansätze, namely dVQE \cite{dvqe1} and the Hermitian-preserving ansatz (HPA)~\cite{dvqe2}, with comparable numbers of variational parameters. The dVQE circuit contains 2312 variational parameters, while the HPA ansatz has 2304 parameters. We define the infidelity with respect to exact diagonalization (ED) as \(1-|\langle\rho|\rho_\text{ED}\rangle|^2\), which directly quantifies the proximity between the density matrix obtained from our algorithm and the exact ED solution.
\begin{figure}[ht]
    \centering
    \includegraphics[width=1\linewidth]{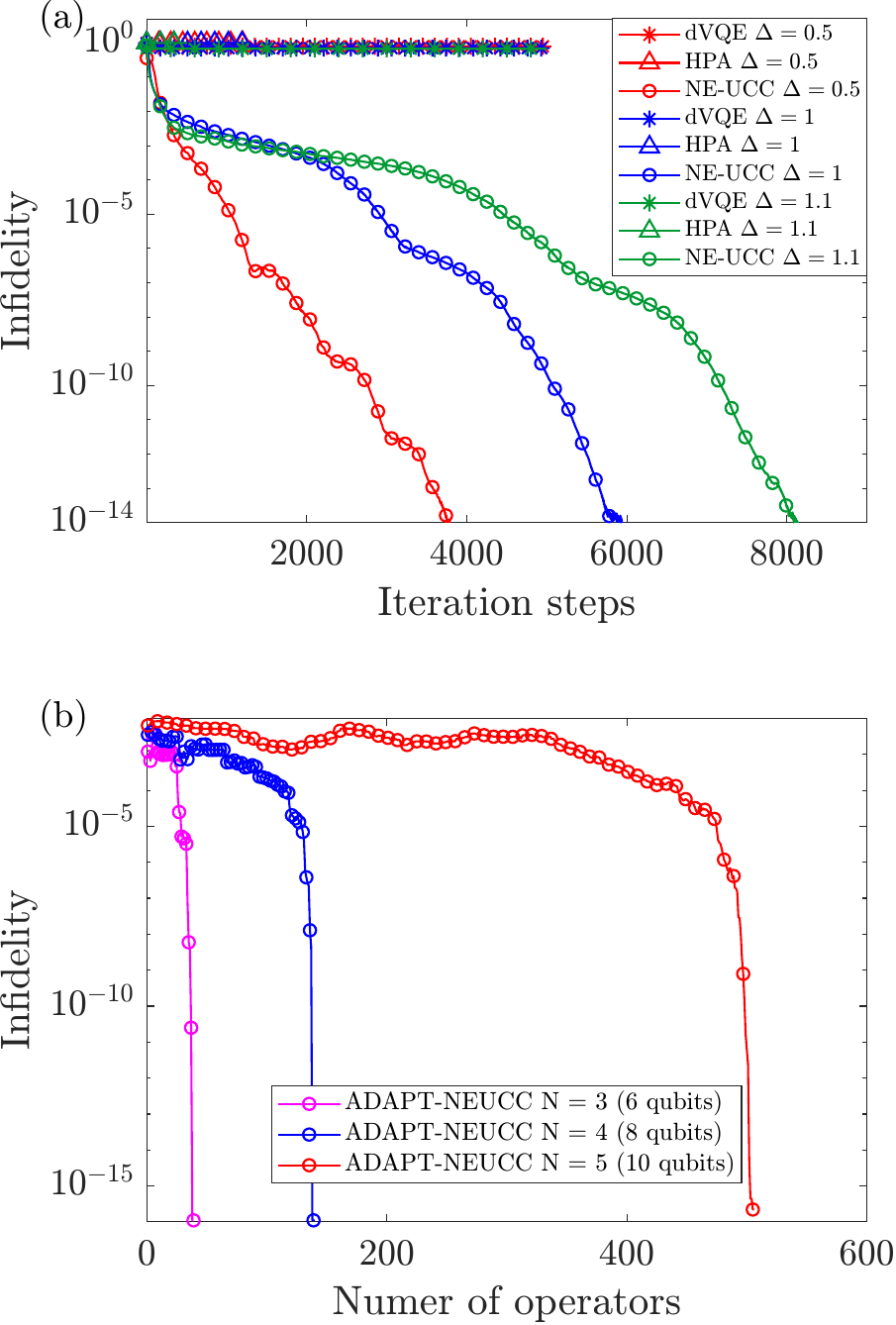}
    \caption{(a) NESS infidelity for an \(n=6\) (12-qubit) XXZ model with \( J=\varepsilon=\mu=1\). The dVQE and HPA circuits have 2312 and 2304 variational parameters, respectively. (b) NESS infidelity as a function of the number of ansatz operators in ADAPT-NE-UCC for XXZ models  with \( J=\mu=1,\Delta=0.5,\varepsilon=0.1\).}
    \label{fig:1}
\end{figure}

\begin{figure}[ht]
    \centering
    \includegraphics[width=0.85\linewidth]{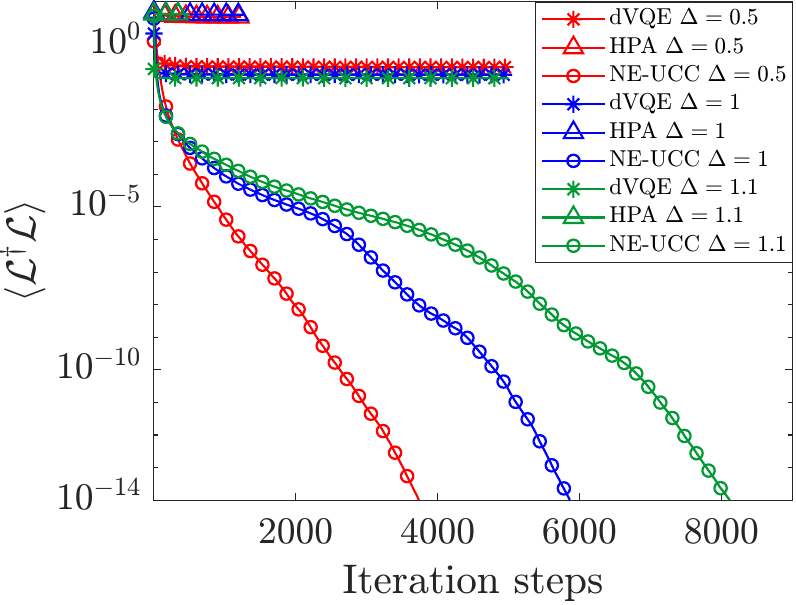}
    \caption{Numerical simulation of the cost function \(\langle \hat{\mathcal{L}}^\dagger\hat{\mathcal{L}}\rangle\) for the $n=6$ (12-qubit) XXZ model with $J=\varepsilon=\mu=1$ and (a) $\Delta=0.5$, (b) $\Delta=1.0$, and (c) $\Delta=1.1$.}
    \label{fig:xxzcostfun}
\end{figure}

As illustrated in Fig.~\ref{fig:1}.~(a), both dVQE and HPA exhibit the barren plateau phenomenon, which hinders the achievement of low infidelity values. In contrast, NE-UCC converges reliably to machine precision, leading to an accuracy difference exceeding ten orders of magnitude. Fig.~\ref{fig:xxzcostfun} displays the decay curves of the cost function $\langle \hat{\mathcal{L}}^\dagger\hat{\mathcal{L}}\rangle$. For NESS, the ideal loss should be close to machine precision. The NE-UCC ansatz converges robustly and reaches a precision as low as \(10^{-15}\), while the other two approaches fail to converge and yield nearly flat decay profiles.

Our approach can be naturally extended to the ADAPT-VQE framework~\cite{vqe2} to balance circuit depth and accuracy. In ADAPT-NEUCC, the density matrix state vector with finite variational parameters is constructed
\begin{align}
    |\rho^{(k+1)}\rangle=e^{\theta_{(k+1)}{t}_{(k+1)}}|\rho^{(k)}\rangle,
\end{align}
where the excitation operators are iteratively taken from the operator pool \(\mathcal{O}\) defined in Eq.(\ref{eq:pool}). In each iteration, ADAPT-VQE employs a standard VQE procedure to minimize the cost function Eq.(\ref{eq:costfun}) and computes the \(L_2\)-norm of gradients \(g_i\) in \(k\)-th iteration with respect to all candidate excitation operators in operator pool,
\begin{align}
    g_i=\bigl|\frac{\partial \mathcal{E}}{\partial \theta_i}\bigr|_{\theta=0}=\bigl|\langle \rho^{(k)}|[{\mathcal{L}}^\dagger{\mathcal{L}},{t}_i]|\rho^{(k)}\rangle\bigr|.
\end{align}
The operator \({t}_i\) with the largest gradient magnitude \(g_i=\max(g)\) is then appended in the current quantum circuit for \((k+1)\)-th iteration. The procedure is repeated until a predefined convergence threshold is satisfied. In Fig.~\ref{fig:1}.~(b), we show the infidelity as a function of circuit depth. The initial point corresponds to the error of the mean-field solution, and subsequent iterations of ADAPT-NE-UCC reduce the infidelity to \(10^{-15}\).
\begin{figure}[h]
    \centering
    \includegraphics[width=0.9\linewidth]{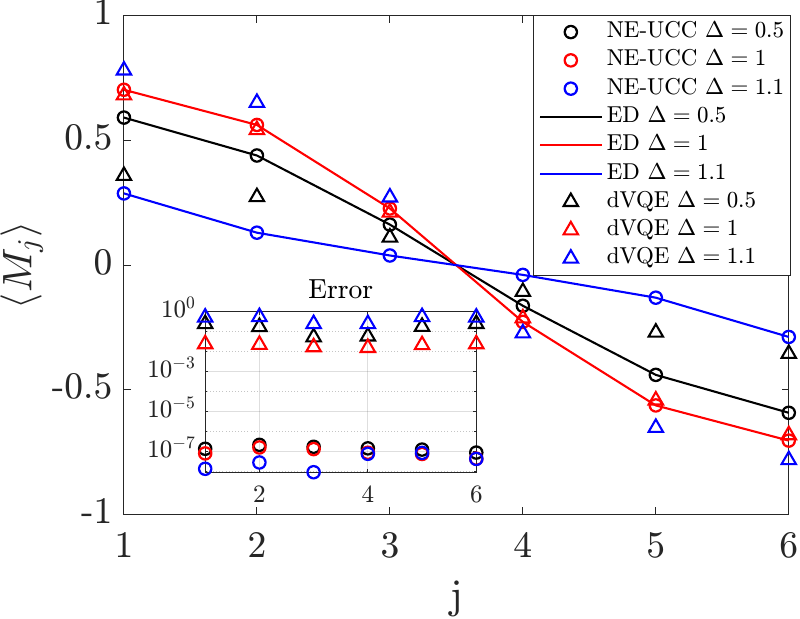}
    \caption{Scaled magnetization \(\langle M_j\rangle \) for strongly driven (\(\mu=1\)) 6-site XXZ model with \( J=\varepsilon=1\). Inset shows the corresponding absolute errors relative to ED result. }
    \label{fig:xxz}
\end{figure}

With the density matrix obtained, we can calculate properties such as the scaled magnetization
\begin{align}
    \langle M_j \rangle=\varepsilon^{-2}\langle I|\sigma_j^z|\rho\rangle /\langle I|\rho\rangle .
\end{align}
As shown in Fig.~\ref{fig:xxz}, very accurate results can be obtained from the NE-UCC density matrix. The corresponding absolute errors relative to the ED benchmark have a magnitude of  \(10^{-7}\). In contrast, dVQE exhibits errors on the order of \(10^{-1}\). 

\begin{table}[h]
    \centering
    \caption{Number of non-zero (nnz) elements in the density matrix obtained from NE-UCC and ED.}
    \begin{ruledtabular}
    \begin{tabular}{ccccc}
        $N$ & $\displaystyle\binom{2N}{N}$ & $\mathrm{nnz}(|\rho_{\scriptscriptstyle \text{NE-UCC}}\rangle)$ & $\mathrm{nnz}(|\rho_{\scriptscriptstyle \text{ED}}\rangle)$ & $\displaystyle\frac{\binom{2N}{N}}{2^{2N}}$ \\
        \hline
        3 & 20 & 20 & 20 & 0.3125 \\
        4 & 70 & 70 & 70 & 0.2734 \\
        5 & 252 & 252 & 252 & 0.2461 \\
        6 & 924 & 924 & 924 & 0.2256 \\
        20 & $\approx 1.38 \times 10^{11}$ & \quad & \quad & $\approx 0.125$ \\
    \end{tabular}
    \end{ruledtabular}
    \label{tab:1}
\end{table}

An important mechanism for such a good performance achieved by NE-UCC and its ADAPT variant is the strong restriction on the search space, which can be understood from the number of non-zero elements in the density matrices. For any vector (bit string) \(|n\rangle \otimes|m\rangle\) within the subspace \({N}_a-{N}_{\tilde{a}} =0\), the first \(N\) qubits \(|n\rangle\) and the last \(N\) qubits \(|m\rangle\) must possess equal Hamming weights; consequently, the maximum dimension of this subspace is \(\binom{2N}{N}\). As shown in Table~\ref{tab:1}, the circuit generated by the NE-UCC ansatz strictly confines the variational manifold to this subspace. Furthermore, the weak symmetry ensures the uniqueness of the steady state~\cite{symmetry1}. 

\subsection{Interacting quantum heat engine}
Our second example is a quantum heat engine with a three-site interacting working medium~\cite{genal3}, described by the system Hamiltonian
\begin{equation}
    \begin{aligned}
    H_{\text{S}}&=\sum_{s=1}^D \varepsilon_s a_s^\dagger a_s+\sum_{s=1}^{D-1}t_s(a_{s+1}^\dagger a_s+ h.c.)\\
    &+\sum_{s=1}^{D-1}Ua_s^\dagger a_s a_{s+1}^\dagger a_{s+1},
\end{aligned}
\end{equation}
where \(D=3\). Taking into account the coupling to left (\(\alpha=L\)) and right (\(\alpha=R\)) thermal reservoirs (leads), the total Hamiltonian reads \(H =H_\text{S}+H_\text{K}+H_\text{SK}\), where
\begin{align}
    H_\text{K}=\sum_{k\alpha}\varepsilon_{k\alpha}a_{k\alpha}^{\dagger}a_{k\alpha}
\end{align}
is the Hamiltonian of the reservoirs, and the system–reservoir coupling is given by
\begin{align}
    H_{\text{SK}}=-\sum_{sk\alpha}(t_{sk\alpha}a_{k\alpha}^{\dagger}a_{s}+h.c.) .
\end{align}
The reservoirs act as particle sources, and the left and right leads are assumed to be in thermodynamic equilibrium at temperatures \(T_\text{L}\) and \(T_\text{R}\) and chemical potentials \(\mu_{\scriptscriptstyle\text{L}}\) and \(\mu_{\scriptscriptstyle\text{R}}\),  respectively. To describe the resulting incoherent particle exchange, we employ two sets of jump operators, 
\begin{align}
    L_{k\alpha}^{-}&=\sqrt{\gamma_{k\alpha}(1-f_{k\alpha})}a_{k\alpha}\\ L_{k\alpha}^{+}&=\sqrt{\gamma_{k\alpha}f_{k\alpha}}a_{k\alpha}^{\dagger}
\end{align}
where \(f_{k\alpha}=[1+\exp((\varepsilon_{k\alpha}-\mu_{\alpha})/T_{\alpha})]^{-1}\) is the Fermi-Dirac distribution function. The dissipation rate is taken to be uniform across all modes, \(\gamma_{k\alpha}=\gamma=2\Delta\varepsilon\), where \(\Delta\varepsilon\) denotes the energy spacing between adjacent levels in each lead, defined as \(\Delta\varepsilon=\frac{E_\text{max}-E_\text{min}}{N-1}\), with \(N\) being the number of discrete energy levels used to discretize the lead spectrum.
\begin{figure}[!ht]
    \centering
    \includegraphics[width=0.9\linewidth]{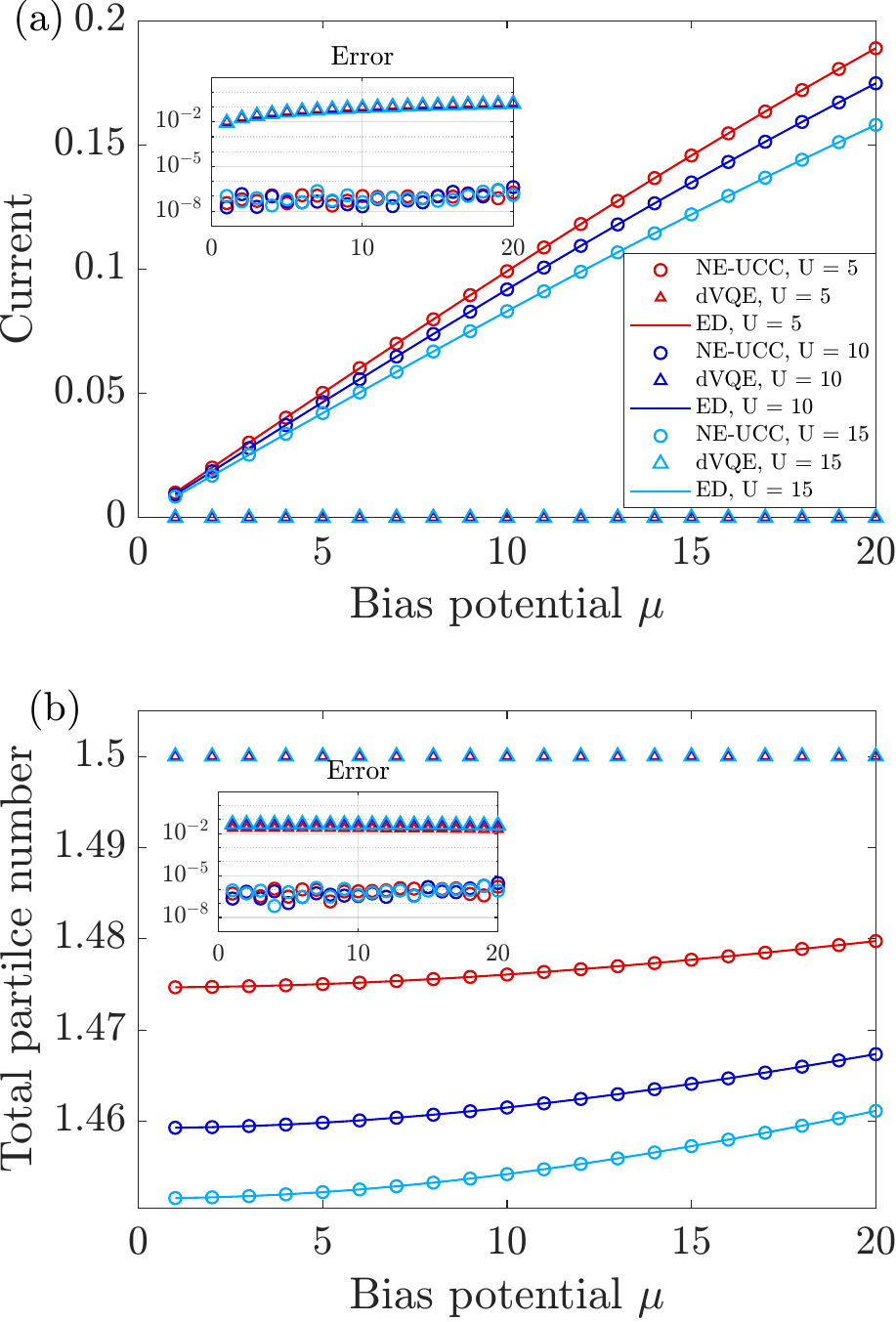}
    \caption{(a) Current and (b) total particle number as functions of the bias potential \(\mu\) for various interaction strengths \(U\). The two leads are maintained at identical temperatures \(T_{\scriptscriptstyle{\text{L}}}=T_{\scriptscriptstyle{\text{R}}}=10\) with symmetric chemical potentials \(\mu_{\scriptscriptstyle{\text{L}}}=-\mu_{\scriptscriptstyle{\text{R}}}=\mu\). The insets display the corresponding absolute deviations from ED.}
    \label{fig:nsite}
\end{figure}

Here, we simulate the non-equilibrium transport with \( N = 2 \) lead levels, totaling 14 qubits. We set on-site energies \(\varepsilon_s\) sampled from a uniform distribution over the interval \([E_\text{min}/5,E_\text{max}/5]\), while the lead levels are uniformly distributed over \([E_\text{min},E_\text{max}]\), with \(t_s=5/8\), \(t_{sk\alpha}=\sqrt{\frac{1}{2\pi\eta}}\), \(\eta=N/(E_\text{max}-E_\text{min})\), and we choose \(E_\text{max}=-E_\text{min}=5\). We compute the steady-state current
$$\langle J_{\alpha}\rangle=\frac{\langle I|-i \sum_{sk}t_{sk\alpha}(a_{k\alpha}^{\dagger}a_{s}-a_{s}^{\dagger}a_{k\alpha})|\rho\rangle}{\langle I|\rho\rangle}$$
as a function of the bias potential \(\mu\), as presented in Fig.~\ref{fig:nsite}.~(a). Fig~\ref{fig:nsite}.~(b) shows the the total site occupation
\begin{align}
    \langle N_{\text{total}}\rangle=\sum_{s=1}^3\frac{\langle I| a_s^\dagger a_s|\rho\rangle}{\langle I|\rho\rangle}
\end{align}
varying with the bias voltage. The steady-state expectation values of these observables exhibit excellent agreement with the exact diagonalization (ED) benchmarks. For a fair comparison, we adopt around 4500 variational parameters in the dVQE ansatz to match the parameter scale of the NE-UCC scheme. At this parameter scale, the hardware-efficient ansatz suffers from severe gradient vanishing issues. Consequently, dVQE fails to qualitatively reproduce the variation trends of the physical observables. In sharp contrast, the NE-UCC approach maintains effective trainability and reliable optimization performance.

\subsection{Variational searching for excited eigenmodes}
Beyond accurately capturing NESS, NE-UCC can also access excited eigenmodes. In general, the eigenvalues of the Liouvillian satisfy
\begin{align}
0 = \lambda_0 \ge \operatorname{Re}\lambda_1 \ge \operatorname{Re}\lambda_2 \ge \cdots,
\end{align}
where \(\lambda_0=0\) corresponds to the steady state. The Liouvillian gap~\cite{open,lgap} \(\Delta \mathcal{L}\) is then defined as
\begin{align}
\Delta \mathcal{L} = \bigl|\operatorname{Re}\lambda_1\bigr|,
\end{align}
which characterizes the separation between the largest real part (equal to zero, associated with the steady state) and the second-largest real part among all Liouvillian eigenvalues. This fundamental quantity governs the relaxation time to the steady state and characterizes dissipative phase transitions~\cite{lgap2,lgap3,lgap4} and exotic chiral damping~\cite{lgap5}

To search for excited eigenmodes near the trial value \(E_0\), we can replace the original cost function \(\mathcal{E} = \min_{\vec{\theta}} \langle \rho(\vec{\theta})|{\hat{\mathcal{L}}}^\dagger{\hat{\mathcal{L}}}|\rho(\vec{\theta})\rangle\) with the energy variance 
\begin{equation}
    \begin{aligned}
    \mathcal{E} = \min_{\vec{\theta},\,\operatorname{Re}E_0,\,\operatorname{Im}E_0}
\langle \rho(\vec{\theta})|
&\bigl(-i\hat{\mathcal{L}}-E_0\bigr)^\dagger \bigl(-i\hat{\mathcal{L}}-E_0\bigr)
|\rho(\vec{\theta})\rangle\\
&+ \kappa\,\bigl|\langle I|\rho(\vec{\theta})\rangle\bigr|^2
\label{eq:excost}
\end{aligned}
\end{equation}
as proposed in Ref.~\cite{cplex}. The penalty term \(\kappa|\langle I|\rho(\vec{\theta})\rangle|^2\) serves to exclude the steady state but it distorts the natural gradient structure of the cost function. Notably, the Liouville operator adopted in this paper differs by a factor of \(-i\) from those used in other literature. 
\begin{figure}[!ht]
    \centering
    \includegraphics[width=1\linewidth]{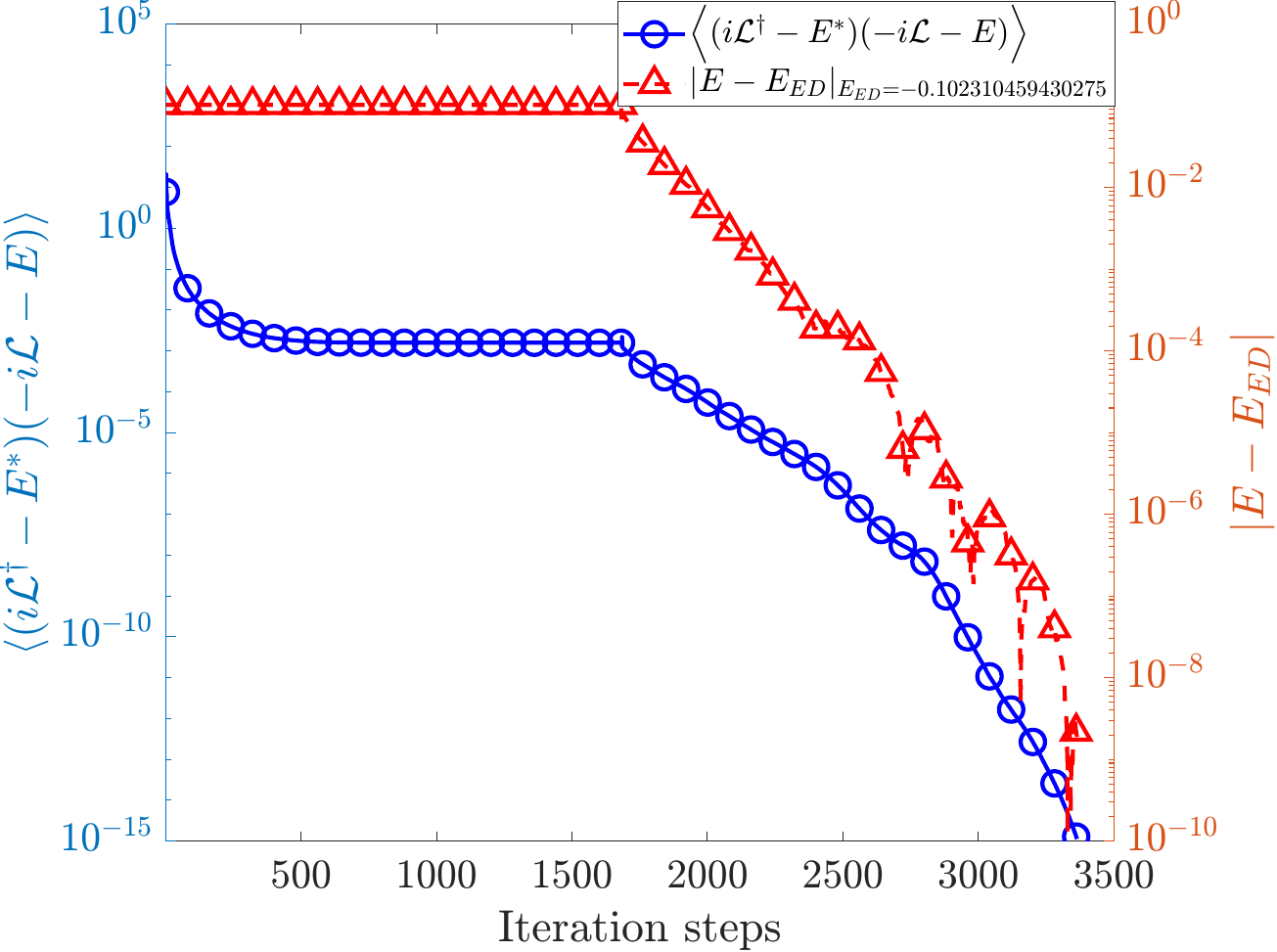}
    \caption{Liouvillian gap optimization for the \(n=6\) XXZ model in the subspace \(N_a - N_{\tilde{a}}=\pm 1\) with  \( J=\Delta=\varepsilon=\mu=1\).}
    \label{fig:ex}
\end{figure}
Since NE-UCC conserves the particle-number difference, the optimization subspace is uniquely fixed simply by choosing an initial state that respects this symmetry. For instance, for excited modes corresponding to \(N_a - N_{\tilde{a}} = \pm k\) (where \(k\) is an integer), it suffices to construct an initial state of the form \(\sum_{m,n}|m\rangle|n\rangle\) satisfying \(\operatorname{HW}(|m\rangle) - \operatorname{HW}(|n\rangle) = \pm k\), where \(\operatorname{HW}(\cdot)\) denotes the Hamming weight. The penalty term can then be naturally eliminated when the excited modes reside in subspaces with \(N_a - N_{\tilde{a}} \neq 0\).
\begin{figure*}[!ht]
    \centering
    \includegraphics[width=0.87\linewidth]{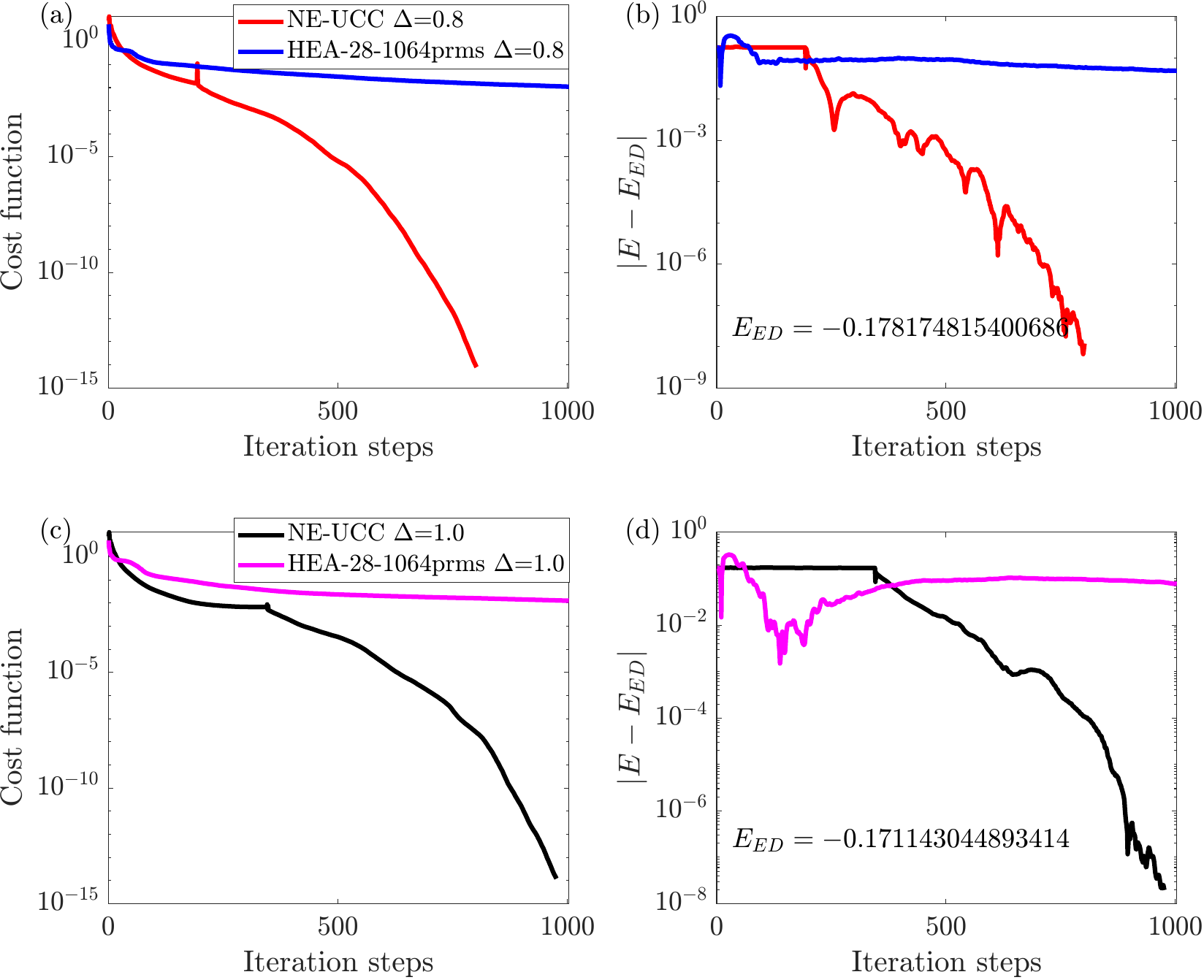}
    \caption{Comparison of Liouvillian gap optimization between NE-UCC and HEA circuit for the \(n=5\) (10-qubit) XXZ model. (a) and (c) show the optimization curves of the cost function for \(\Delta=0.8\) and \(\Delta=1.0\), respectively, while (b) and (d) present the absolute Liouvillian gap errors.}
    \label{fig:compare}
\end{figure*}

In Fig.~\ref{fig:ex}, we present the optimization process of the excited modes for the \(n=6\) (12-qubit) XXZ model with \(J = \Delta = \varepsilon = \mu = 1\) in the corresponding subspace \(N_a - N_{\tilde{a}} = \pm 1\). Its Liouvillian gap lies exactly within this subspace, corresponding to a degenerate eigenstate. The plateau region in the figure corresponds to the pre-training stage with fixed parameter \(E_0\), after which \(E_0\) is released for full subsequent optimization. Our results clearly demonstrate that NE-UCC can achieve high precision (better than \(10^{-8}\)) for the Liouvillian gap. Details of the algorithm are provided in the Appendix.\ref{ag:ex}.

In contrast, HEA-based circuits~\cite{cplex} suffer from optimization bottlenecks as shown in Fig.~\ref{fig:compare}. Here, we perform a comparative analysis for the \(n=5\) XXZ model, with parameters \(J =\varepsilon = \mu = 1\), between the HEA circuit from Ref~\cite{cplex} (28 layers with 1064 parameters) and the NE-UCC ansatz with 1050 parameters, whose parameter counts are well matched. Identical initial states are adopted for both the HEA and NE-UCC ansatz throughout the numerical tests. The HEA ansatz exhibits prominent barren plateaus at moderate circuit depth, accompanied by extremely slow convergence and a limited final accuracy on the order of \(10^{-1}\). In contrast, NE-UCC maintains an extremely high convergence accuracy throughout the optimization, with all gap errors below \(10^{-8}\). The above results demonstrate that NE-UCC outperforms HEA remarkably in both convergence speed and computational accuracy.

\section{Conclusion}
To summarize, we have established a variational implementation of the non-equilibrium unitary coupled cluster ansatz for open fermionic systems, which utilizes a weak symmetry arising from the Liouvillian dynamics that conserves the particle-number difference. This symmetry acts as a safeguard, restricting the variational search to a small subspace and ensuring physical consistency. Consequently, the NE-UCC approach delivers highly accurate results (infidelity $<10^{-10}$) for both steady states and excited modes in strongly correlated regimes, significantly exceeding the accuracy of prior works. This physically motivated ansatz thus offers a vital pathway to design high-performance variational quantum circuits. Future work can focus on generalizing the environment to include memory effects, utilizing techniques such as pseudomode Lindblad methods~\cite{plind1,plind2} to capture structured environments and memory-driven dynamics.

\begin{acknowledgments}
The authors would like to thank Professors Alan A. Dzhioev and D. S. Kosov for their helpful discussions on superfermion theory. This work was supported by NSFC (22393913), by the Strategic Priority Research Program (XDB0450101) and the robotic AI-Scientist platform of the Chinese Academy of Sciences, and by USTC Supercomputer Center.
\end{acknowledgments}

\appendix
\section{Analytical expressions of superfermion approaches}
\label{apd:sfscf}
We consider a simple illustrative example featuring a quadratic Hamiltonian \(\bm{H}=\bm{E}\bm{a}^\dagger \bm{b}\), together with jump operators \(\bm{L}^{(1)}=\sqrt{\bm{\Gamma}^-} \bm{a}\) and \(\bm{L}^{(2)}=\sqrt{\bm{\Gamma}^+} \bm{a}^\dagger\) (with tilde form counterparts such as \(\tilde{\bm{H}}=\bm E \tilde{\bm a}^\dagger \tilde{\bm b}\)). Here \(\bm{E}\) denotes a Hermitian matrix, while \(\bm{\Gamma}\) is a real positive diagonal matrix. The labels \(a\) and \(b\) are introduced to differentiate distinct subscript indices for identical creation and annihilation operators; concretely, \(\bm{E} \bm{a}^\dagger  \bm{b} \sim\sum_{ij} E_{ij}\,a_i^\dagger a_j\) and \(\bm{E}\bm{a}^\dagger \bm{a} \sim\sum_i E_{ii}\, a_i^\dagger a_i\). The Liouvillian within the superfermion formalism then reads
\begin{equation}
    \begin{aligned}
    \hat{\mathcal{L}}&={H}-\tilde{H}-i\sum_k \Pi_k\\
    &=\bm{E( a^\dagger b +\tilde{b}\tilde{a}^\dagger )}-i\bm{(\Gamma^- -\Gamma^+)(a^\dagger a-\tilde{a} \tilde{a}^\dagger)}\\
    &-2\bm{(\Gamma^-\tilde{a}a-\Gamma^+a^\dagger\tilde{a}^\dagger )}- \operatorname{Tr}[\bm{E}+i\bm{(\Gamma^++\Gamma^-)}].
    \end{aligned}
\end{equation}
For quadratic Liouvillian, we can introduce a set of Nambu spinors \(\bm{d}=\begin{pmatrix}\bm{a},\tilde{\bm{a}}^\dagger\\ \end{pmatrix}\), the Liouvillian then is transformed into a Hamiltonian of Bogoliubov–de Gennes (BdG) form~\cite{genal1},
\begin{equation}
    \begin{aligned}
    \hat{\mathcal{L}}&=\bm{d}^\dagger \bm{L} \bm{d}-\operatorname{Tr}(\bm{E}+i\bm{(\Gamma^++\Gamma^-)}),
\end{aligned}
\end{equation}
where \begin{equation}
    \begin{aligned}
        \bm{L}=\begin{pmatrix}
        \bm{E}-i\bm{(\Gamma^--\Gamma^+)}&2\bm{\Gamma^+}\\
        -2\bm{\Gamma^-}&\bm{E}+i\bm{(\Gamma^--\Gamma^+)}\\
    \end{pmatrix}.
    \end{aligned}
    \label{eq:BDG}
\end{equation}
The Liouvillian can be diagonalized as
\begin{align}
    \hat{\mathcal{L}}= \bm{\bar{\xi}}\bm{\varepsilon}\bm{\xi}-\bm{\eta},
\end{align}
here \(\bm{\varepsilon}=\bm{V}^{-1} \bm{L} \bm{V}\) is a complex diagonal matrix with pairwise complex conjugate eigenvalues and we get some \(\bm{\xi}\)-modes that
\begin{align}
    \bm{\bar{\xi}}=\bm{d}^\dagger \bm{V}
    \label{eq:D1}
\end{align} 
and
\begin{align}
    \bm{\xi}=\bm{V}^{-1}\bm{d}.
    \label{eq:D2}
\end{align}
The complex (non-canonical) Bogoliubov transformation preserves the canonical anti-commutation relations,
\begin{align}
    \{\xi_i,\bar{\xi}_j\}=\delta_{ij},
    \label{eq:anti}
\end{align}
but \(\xi_i^\dagger \ne \bar{\xi}_i\), since \(\bm{V}\) is not unitary. We can immediately obtain the equations of motion:
\begin{align}
    [\hat{\mathcal{L}},\xi_i]&=-\varepsilon_i\xi_i,\\
    [\hat{\mathcal{L}},\bar{\xi}_i]&=\varepsilon_i \bar{\xi}_i.
\end{align}
and the time evolution of \(\bm \xi\) operators is
\begin{equation}
    \begin{aligned}
    \xi_i(t)&=e^{-i\hat{\mathcal{L}}t}\xi_i e^{i\hat{\mathcal{L}}t}\\
    &=\xi_i-it[\hat{\mathcal{L}},\xi_i]-\frac{t^2}{2!}[\hat{\mathcal{L}},[\hat{\mathcal{L}},\xi_i]]\\
    &+\frac{it^3}{3!}[\hat{\mathcal{L}},[\hat{\mathcal{L}},[\hat{\mathcal{L}},\xi_i]]]+\dots \\
    &=\xi_i+it\varepsilon_i\xi_i-\frac{t^2}{2!}\varepsilon_i^2\xi_i-\frac{it^3}{3!}\varepsilon_i^3\xi_i+\dots\\
    &=e^{i\varepsilon_i t}\xi_i,
\end{aligned}
\end{equation}
and similarly
\begin{align}
    \bar{\xi}_i(t)&=e^{-i\varepsilon_i t}\bar{\xi}_i.
\end{align}
Noting that the time evolution of density matrix in the Schrödinger picture is opposite to that of the operators in the Heisenberg picture.
\subsection{Spectrum of the noninteracting Liouvillian and non-equilibrium steady state}
For the steady state with \(t \rightarrow{\infty}\), it follows that
\begin{align}
    \hat{\mathcal{L}}|\rho_{\infty}\rangle=0.
\end{align}
Considering the expectation value of the form
\begin{equation}
    \begin{aligned}
    \langle I|\bar{\xi}_i(t)\xi_j(t)&|\rho_{\infty}\rangle\\
    &=e^{i\operatorname{Re}(\varepsilon_j-\varepsilon_i)t}e^{-\operatorname{Im}(\varepsilon_j-\varepsilon_i)t}\langle I|\bar{\xi}_i\xi_j|\rho_{\infty}\rangle ,\\
   &=e^{i\operatorname{Re}(\varepsilon_j-\varepsilon_i)t}e^{-\operatorname{Im}(\varepsilon_j-\varepsilon_i)t}\langle I|\delta_{ij}-\xi_j\bar{\xi}_i|\rho_{\infty}\rangle
\end{aligned}
\end{equation}
the amplitude \(e^{-\operatorname{Im}(\varepsilon_k) t}\) diverges if \(\operatorname{Im}(\varepsilon_k)<0\), and \(e^{\operatorname{Im}(\varepsilon_k) t}\) diverges if \(\operatorname{Im}(\varepsilon_k)>0\), so we must have
\begin{align}
    \xi_k|\rho_{\infty}\rangle=0, \quad \text{for}\quad\operatorname{Im}(\varepsilon_k)<0
        \label{eq:s1}
    \end{align}
\begin{align}
    \bar{\xi}_k|\rho_{\infty}\rangle=0, \quad \text{for}\quad\operatorname{Im}(\varepsilon_k)>0
\label{eq:s2}
\end{align}
Similarly to keep \(\langle I|\xi_i(t)\bar{\xi}_j(t)|\rho_{\infty}\rangle \) finite, we have
\begin{align}
    \langle I| \xi_k&=0, \quad \text{for}\quad\operatorname{Im}(\varepsilon_k)>0
    \label{eq:s3}
\end{align}
\begin{align}
    \langle I| \bar{\xi}_k&=0, \quad \text{for}\quad\operatorname{Im}(\varepsilon_k)<0.
    \label{eq:s4}
\end{align}
By reordering and pairing \(\bm\xi\) \((\bm{\bar\xi})\), we can obtain the creation (\(\bm c^\dagger\), \(\tilde{\bm c}^\dagger\)) and annihilation operators (\(\bm c\), \(\tilde{ \bm c}\)) of quasiparticles as well as the corresponding left and right quasiparticle vacuum, that is
\begin{equation}
    \begin{aligned}
    \langle I|c_i^\dagger &=0,\quad \langle I|\tilde{c}_i^\dagger =0,\\
    c_i|\rho_{\infty}\rangle&=0,\quad \tilde{c}_i|\rho_{\infty}\rangle=0.
    \label{eq:nqp}
\end{aligned}
\end{equation}
The noninteracting (quadratic) Liouvillian then gives the complex spectrum 
\begin{align}
    \hat{\mathcal{L}}=\sum_i^{\dim(\bm{\xi})/2} \varepsilon_i c_i^\dagger c_i -  \varepsilon_i^* \tilde{c}_i^\dagger \tilde{c}_i,
\end{align}
but \(c_i\) and \(c_i^\dagger\) (\(\tilde{c}_i\) and \(\tilde{c}_i^\dagger\)) are not Hermitian conjugated to each other. 

In this paper, the steady-state solution of the superfermion mean-field (MF) approach is used as an optional initial guess for variational quantum computation. To this end, the steady-state solution \(|\rho_{0}\rangle\) needs to be mapped onto the computational basis. Here, we directly present the computational formula for \(|\rho_{0}\rangle\). In fact, we can always select the product operators \(\prod_{i=1}^{n} c_i\tilde{c}_i \) acting on a certain state \(|\Psi\rangle\), i.e., 
\begin{align}
    |\rho_{0}\rangle=\prod_{i=1}^{n} c_i\tilde{c}_i |\Psi\rangle.
    \label{eq:v0}
\end{align}
such that any quasi-particle annihilation operator acting on the vacuum state yields zero, because
\begin{align}
    c_m\prod_n c_n \tilde{c}_n =(-1)^{2(m-1)}(\prod_{n\neq m}c_n \tilde{c}_n)c_m c_m \tilde{c}_m=0,
\end{align}
likewise, \(\tilde{c}_m\prod_n c_n \tilde{c}_n =0 \).
Meanwhile, according to the previous discussion, \( |\rho_{0}\rangle\) must be the eigenstate of \(({N}_a-{N}_{\tilde{a}}) |\rho_0\rangle=0\), and utilizing the fact that \([{N}_a-{N}_{\tilde{a}} ,\prod_n c_n\tilde{c}_n]=0\), we have 
\begin{align}
    ({N}_a-{N}_{\tilde{a}})|\rho_{0}\rangle=\prod_{i=1}^{n} c_i\tilde{c}_i ({N}_a-{N}_{\tilde{a}})|\Psi\rangle=0.
\end{align}
It follows that \(|\Psi\rangle\) is any eigenstate of \({N}_a-{N}_{\tilde{a}}\) with eigenvalue 0. A convenient selection for \(|\Psi\rangle\) is the zero state \(|0\rangle\otimes|\tilde{0}\rangle\).
Since the quasi-particle operators are actually linear combinations of original fermionic operators, as shown in Eqs.~(\ref{eq:D1}) and ~(\ref{eq:D2}), an explicit form of \(|\rho_{0}\rangle\) can be obtained.
\subsection{Expectation values of observables}
After obtaining the transformation coefficient \(\bm V\) by diagonalizing Eq.(\ref{eq:BDG}), physical observables \(O(\bm a^\dagger,\bm a)\) can be directly obtained through algebraic operations of quasiparticles.
Utilizing the anti-commutation relations Eq.(\ref{eq:anti}) along with Eq.(\ref{eq:s1})-Eq.(\ref{eq:s4}), we can rapidly compute the expectation values
\begin{align}
    \langle I|\bar{\xi}_i\xi_j|\rho_{0}\rangle =D_{ij}=\delta_{ij} \Theta[\operatorname{Im}(\varepsilon_i)]
\end{align}
and
\begin{align}
    \langle I|\xi_i\bar{\xi}_j|\rho_{0}\rangle=\bar{D}_{ij}=\delta_{ij}-D_{ij},
\end{align}
with \(\Theta(x)\) the Heaviside step function.
Then, the expectation value of observables such as occupation number \( n_i=\langle a_i^\dagger a_i\rangle \) where 
 \(a_i \in \bm d\) can be computed from the following expression:
 \begin{equation}
     \begin{aligned}
    \langle I|d^\dagger_i d_j|\rho_0\rangle &= \langle I|\sum_{\mu,v}\bar\xi_{\mu} V^{-1}_{\mu,i} V_{j,v}\xi_v|\rho_0\rangle\\
    &=\sum_{\mu,v}\langle I|\bar{\xi}_{\mu} \xi_v|\rho_0\rangle V^{-1}_{\mu,i} V_{j,v}\\
    &=\sum_{\mu,v}D_{\mu,v} V^{-1}_{\mu,i} V_{j,v}\\
    &=[(\bm V^{-1})^{T}\bm D \bm V^{T}]_{i,j}
\end{aligned}
\label{eq:ob1}
 \end{equation}
\begin{equation}
    \begin{aligned}
     \langle I|d_i d^\dagger_j|\rho_0\rangle &=
     \sum_{\mu,v}\langle I|\xi_{\mu} \bar{\xi}_v|\rho_0\rangle V_{i,\mu} V^{-1}_{v,j}\\
     &=\sum_{\mu,v} \bar{D}_{\mu,v}V_{i,\mu} V^{-1}_{v,j}\\
     &=(\bm V\bar{\bm D}\bm V^{-1})_{i,j}
\end{aligned}
\end{equation}

\section{Numerical Simulation of Boundary-Driven XXZ Model}
\label{apd:xxzmd}
Using the Jordan–Wigner transformation
\begin{align}
    {a}_k^\dagger = {\sigma}_z^{\otimes k-1} \otimes {\sigma}_k^+ \otimes \mathds{1}^{\otimes n-k}, \\
    {a}_k = {\sigma}_z^{\otimes k-1} \otimes {\sigma}_k^- \otimes \mathds{1}^{\otimes n-k}
\end{align}
to map Pauli operators onto fermionic operators, the Liouvillian for the boundary-driven XXZ chain in the superfermion representation is given by
\begin{align}
    \hat{\mathcal{L}}&={\mathcal{H}}+{\mathcal{V}},
\end{align}
where
\begin{equation}
    \begin{aligned}
    {\mathcal{H}}&=\sum_i^{n-1}2J(a_i^\dagger a_{i+1}+\tilde{a}_{i+1}\tilde{a}_i^\dagger+a_{i+1}^\dagger a_{i}+\tilde{a}_{i}\tilde{a}_{i+1}^\dagger)\\
        &-2\Delta\sum_i^{n-1}(a_i^\dagger a_i +\tilde{a}_i\tilde{a}_i^\dagger+a_{i+1}^\dagger a_{i+1} +\tilde{a}_{i+1}\tilde{a}_{i+1}^\dagger)\\
        &-i[-(\gamma^--\gamma^+)a_1^\dagger a_1+ (\gamma^- -\gamma^+)\tilde{a}_1\tilde{a}_1^\dagger \\
        &-2i\gamma^+\tilde{a}_1a_1+2i\gamma^-a_1^\dagger \tilde{a}_1^\dagger ]\\
        &-i[(\gamma^--\gamma^+)a_n^\dagger a_n- (\gamma^- -\gamma^+)\tilde{a}_n\tilde{a}_n^\dagger \\
        &-2i\gamma^-\mathcal{S}_n\tilde{\mathcal{S}}_n\tilde{a}_n a_n+2i\gamma^+ \mathcal{S}_n\tilde{\mathcal{S}}_n a_n^\dagger \tilde{a}_n^\dagger ]\\
        &-2i(\gamma^-+\gamma^+)+4(n-1)\Delta,
        \label{eq:H}
\end{aligned}
\end{equation}
\begin{align}
{\mathcal{V}}=4\Delta\sum_i^{n-1}( {n}_i{n}_{i+1}-\tilde{n}_i\tilde{n}_{i+1}),
\end{align}
and \(\mathcal{S}_n=\prod_{k=1}^{n-1} \sigma_k^z=\prod_{k=1}^{n-1}(1-2a_k^\dagger a_k)\) is the string operator introduced by the JW transformation. Referring to our previous convention, \(\tilde{a}_i=a_{i+n}\), and thus \(\tilde{\mathcal{S}}_n=\prod_{k=1}^{n-1} \sigma_{k+n}^z=\prod_{k=1}^{n-1}(1-2\tilde{a}_{k}^\dagger \tilde{a}_{k})\). It is straightforward to verify via Eq.(\ref{eq:adj0}) that \(\mathcal{S}_n|I\rangle=\tilde{\mathcal{S}}_n|I\rangle\). Furthermore, since \(\tilde{\mathcal{S}}_n\) commutes with the original density matrix, it appears in Eq.(\ref{eq:H}). Notably, \(\mathcal{S}_1=\tilde{\mathcal{S}}_1=1\) and hence they are omitted.
To provide a suitable initial state \(|\rho_0\rangle\) for quantum computations, we decouple the higher-order interactions \({\mathcal{V}}\) via the Hartree-Fock (HF) mean-field approximation,
\begin{equation}
    \begin{aligned}
        {V}_0&=\operatorname{HF}({\mathcal{V}})\\
&=4\Delta\sum_i^{n-1}(\langle a_{i+1}^\dagger a_{i+1} \rangle a_{i}^\dagger a_{i}+\langle a_{i}^\dagger a_{i} \rangle a_{i+1}^\dagger a_{i+1} \\
&+\langle \tilde{a}_{i+1}^\dagger \tilde{a}_{i+1} \rangle \tilde{a}_{i}\tilde{a}_{i}^\dagger +\langle \tilde{a}_{i}^\dagger \tilde{a}_{i} \rangle \tilde{a}_{i+1}\tilde{a}_{i+1}^\dagger )\\&-4\Delta\sum_i^{n-1}(\langle a_{i+1}^\dagger a_{i+1} \rangle +\langle {a}_{i}^\dagger {a}_{i} \rangle).
    \end{aligned}
\end{equation}
Simultaneously, we define \( H_0\) by omitting the string operators in the Hamiltonian \({\mathcal{H}}\) i.e., approximating \(\mathcal{S}_n=\tilde{\mathcal{S}}_n=1\).
Consequently, we obtain the zeroth-order Liouvillian \(\hat{\mathcal{L}}_0\) and corresponding steady state \(|\rho_0\rangle\) under the mean-field (MF) approximation,
\begin{align}
    \hat{\mathcal{L}}_0|\rho_0\rangle=( H_0+ V_0)|\rho_0\rangle=0.
    \label{eq:scf}
\end{align}
It can be further expressed as a operator matrix in vector notation with auxiliary vector \( \bm{d}=\begin{pmatrix}
    a_1,\dots,a_n,\tilde{a}_1^\dagger,\dots,\tilde{a}_n^\dagger
\end{pmatrix}\),
\begin{equation}
    \begin{aligned}
     \hat{\mathcal{L}}_0&=\bm{d}^\dagger \begin{pmatrix}
     \bm{H+U}-i\bm{\Omega} & 2\bm{\Gamma}^+ \\
     -2\bm{\Gamma}^- & \bm{H+U}+i\bm{\Omega}
     \end{pmatrix} \bm{d} \\
     & \quad-\bm{Tr}(\bm{H+U}+i\bm{\Lambda})\\
     &=\bm{d}^\dagger \bm{V}^{-1} \bm{\varepsilon} \bm{V}\bm{d}+c
     \label{eq:scf2}
\end{aligned}
\end{equation}
where 
\begin{align}
    \bm{H}=\begin{pmatrix}
        -2\Delta&2J&0 &\dots&0\\
        2J &-4\Delta&2J& &\vdots\\
        0  &\ddots& \ddots&\ddots&0\\
        \vdots  & &2J &-4\Delta&2J\\
        0&\dots&\dots&2J&-2\Delta
    \end{pmatrix}
\end{align}
\begin{equation}
    \begin{aligned}
    \bm{U}=4\Delta\operatorname{diag}(&\langle {n}_2\rangle,\langle {n}_1\rangle+\langle {n}_3\rangle,\dots,\\
    &\langle {n}_{n-2}\rangle+\langle {n}_n\rangle,\langle {n}_{n-1}\rangle)
\end{aligned}
\end{equation}

\begin{align}
    \bm{\Gamma}^+=\operatorname{diag}(\gamma^-,0,\dots,0,\gamma^+)
\end{align}
\begin{align}
    \bm{\Gamma}^-=\operatorname{diag}(\gamma^+,0,\dots,0,\gamma^-)
\end{align}
\begin{align}
    \bm{\Omega}&=\bm{\Gamma^-} -\bm{\Gamma^+}
\end{align}
\begin{align}
    \bm{\Lambda}&=\bm{\Gamma^-} +\bm{\Gamma^+}
\end{align}

Since \(\bm U\) contains the unknown quantity \(\langle {n}_i\rangle=\langle a_i^\dagger a_i\rangle\), we need to solve Eq.(\ref{eq:scf}), Eq.(\ref{eq:scf2}) and Eq.(\ref{eq:ob1}) self-consistently for the transformation coefficient \(\bm V\) and \(\langle {n}_i\rangle\), and iterate this process repeatedly until convergence is achieved.
After completing the MF calculation, we obtain the non-equilibrium quasiparticles of Eq.(\ref{eq:nqp}) by reordering \(\bm\xi\) \((\bm{\bar\xi})\). In this representation, the non-equilibrium steady state \(|\rho_{0}\rangle\) serves as the natural vacuum for annihilation operators \(\bm c\) and \(\tilde{ \bm c}\) and is obtained from Eq.~(\ref{eq:v0}) by choosing $|\Psi\rangle=|0\rangle \otimes|\tilde{0}\rangle$.
\begin{align}
    |\rho_{0}\rangle=\prod_{i=1}^{n} c_i\tilde{c}_i |0\rangle \otimes |\tilde{0}\rangle,
\end{align}
which allows a direct mapping onto the computational basis of a quantum circuit.
\section{Benchmark for Exact Diagonalization }
We use the exact diagonalization of superoperator matrices in the Pauli representation as the benchmark, and first verify the validity of different superoperator mappings. Since various mapping schemes yield distinct matrices after transformation into the Pauli basis, direct comparison of wave functions is not feasible. We therefore assess the accuracy of these mappings by comparing the observables calculated via exact diagonalization and analytical solutions for each scheme.
Besides the superfermion formalism, the Choi-Jamiolkowski isomorphism \(A\rho B\rightarrow A \otimes B^T|\rho\rangle\)\cite{vec1,vec2,vec3} is another widely adopted technique to matrixize the Liouvillian operator, which has been employed in numerous studies including dVQE. Within this formalism, the Lindblad equation
\begin{equation}
    \begin{aligned}
    \mathcal{L}\rho(t)=&-i[ H,\rho(t)]+\sum_{i}(2 L_i\rho(t)  L_i^\dagger
-\{ L_i^\dagger  L_i,\rho(t)\}),
\end{aligned}
\end{equation}
has Liouvillian superoperator
\begin{equation}
    \begin{aligned}
\hat{\mathcal{L}}_{choi}&=-i(H\otimes \mathbf {1} - \mathbf {1} \otimes H^{T})\\
&+ \sum_{m}2L_{m}\otimes L_{m}^{*}-L_{m}^{\dagger}L_{m}\otimes \mathbf{1} -\mathbf {1} \otimes L_{m}^{T}L_{m}^{*},
\label{eq:choi}
\end{aligned}
\end{equation}
and the corresponding left vacuum state is given by
\begin{align}
    |I\rangle_{choi}=\sum_q|q\rangle \otimes|q\rangle,
\end{align}
where \(|q\rangle\) denotes an arbitrary bit string.

We use the analytical solution of the boundary-driven XXZ model as a reference. In the weak-coupling limit \(\varepsilon \to 0\) with anisotropy \(\Delta=0.5\), the scaled magnetization admits an analytical expression~\cite{xxz2}:\(\lim_{\varepsilon \to 0} M(j)_{ex}=\frac{\mu}{9}(4(\frac{5}{8})^{j-1}-4(\frac{5}{8})^{n-j}-(-8)^{1-n}((-5)^{j-1}-(-5)^{n-j}))\).
\begin{figure}
    \centering
    \includegraphics[width=0.9\linewidth]{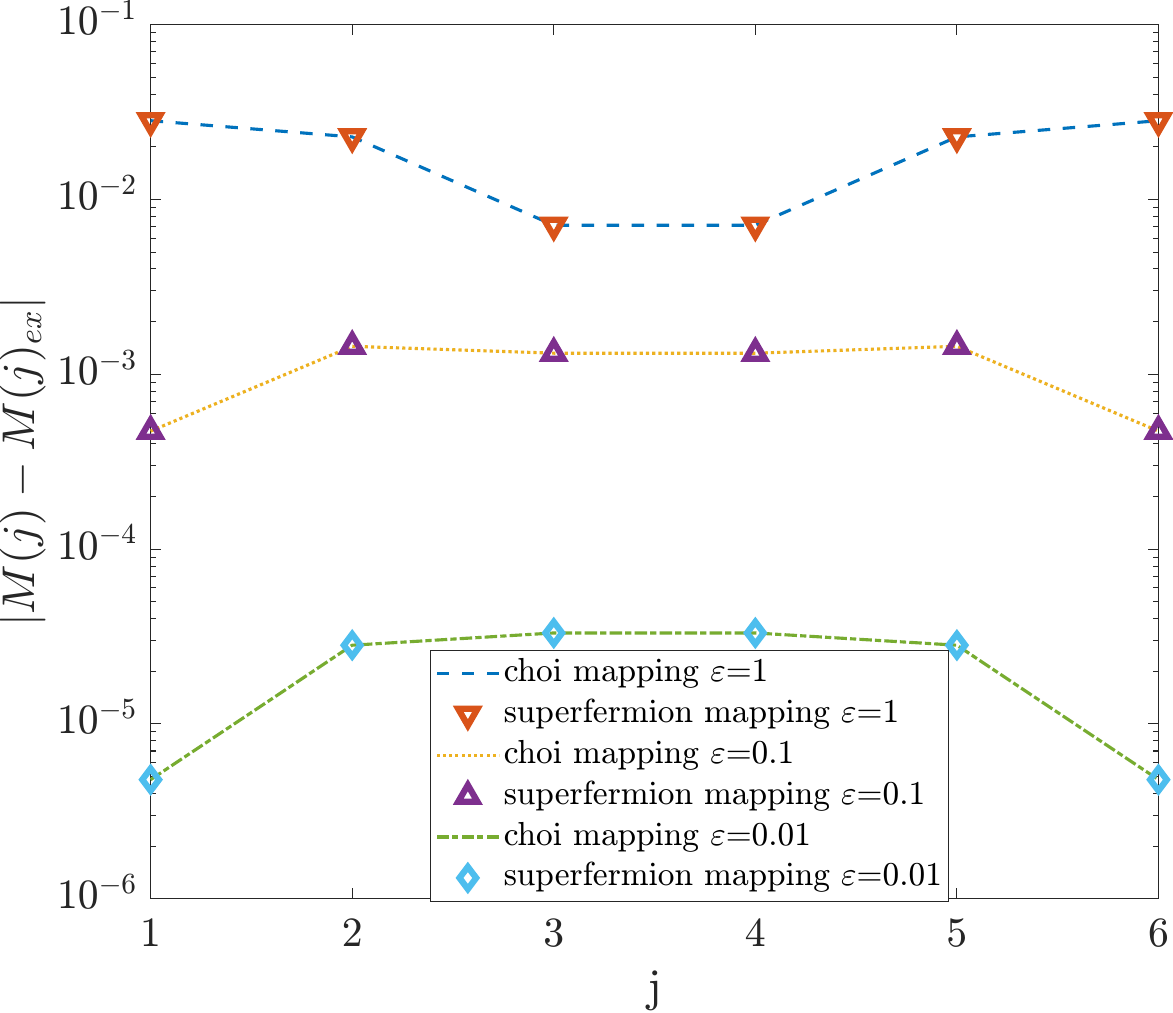}
    \caption{The absolute error of magnetization relative to the asymptotic limit solution for varying coupling strength \(\varepsilon\), with \(n=6\), \(\Delta=0.5\) and \(\mu=0.2\).}
    \label{fig:xxz_bench}
\end{figure}
The Fig.~\ref{fig:xxz_bench} presents the absolute error between the magnetization and the asymptotic limit solution for various values of \(\varepsilon\). As the coupling strength decreases, the results from all mapping schemes converge toward the asymptotic limit with gradually reduced errors. Furthermore, the results obtained via the Choi isomorphism and the superfermion formalism are in agreement throughout. These observations confirm the validity of the adopted mappings. Therefore, exact diagonalization serves as a reliable benchmark.

\section{Variational Search Scheme for Liouvillian Spectra}
We adopt the Liouvillian-gap searching algorithm proposed in Ref~\cite{cplex} to search the excited modes. The core idea of this algorithm is to first fix the initial excitation energy \(E\) and pre-train the gate parameters, thereby driving the trial solution closer to the target excited state. Subsequently, \(E\) is relaxed and trained jointly with the gate parameters until the convergence threshold is reached. In literature~\cite{cplex}, since the HEA cannot guarantee the subspace of the variational manifold, an additional orthogonality penalty term \(\kappa\,\bigl|\langle I|\rho(\vec{\theta})\rangle\bigr|^2\) needs to be imposed during pre-training to ensure distance from the NESS. In the NE-UCC method, by controlling the subspace to which the initial state belongs (specifically the particle number difference), the variational manifold is guaranteed to remain within that subspace. Consequently, for excitation modes that do not lie in the same subspace as the NESS, there is no need to introduce an additional orthogonality penalty term.

\begin{center}
\label{ag:ex}
\rule{0.92\linewidth}{0.4pt}\\[4pt]
\textbf{Algorithm: VQA Search for Liouvillian Gap}\\[4pt]
\rule{0.92\linewidth}{0.4pt}\\[6pt]
\begin{minipage}{0.92\linewidth}
\noindent
1.\quad Initialize the energy parameter \(E=E_0\) and penalty factor \(\kappa=n^2\)~\cite{cplex}\\[4pt]
2.\quad Fix \(\operatorname{Re}E\), optimize \(\operatorname{Im}E\) and \(\vec{\theta}\) to minimize \(\mathcal{E}(\vec{\theta},\operatorname{Im}E)\) defined in Eq.~(\ref{eq:excost})\\[4pt]
3.\quad Remove the constraint on \(\operatorname{Re}E\), optimize \(\operatorname{Re}E\), \(\operatorname{Im}E\), and \(\vec{\theta}\) simultaneously with \(\kappa=0\) to minimize \(\mathcal{E}(\vec{\theta},\operatorname{Re}E,\operatorname{Im}E)\).\\[4pt]
4.\quad \textbf{Return} the optimized eigenvalue \(E\)
\end{minipage}\\[6pt]
\rule{0.92\linewidth}{0.4pt}
\end{center}

\bibliography{reference}

\end{document}